\documentclass[preprint,fleqn,showpacs,showkeys]{revtex4}
\usepackage{graphicx}
\usepackage{amssymb}
\usepackage{amsmath}
\usepackage{bm}
\begin{document}

\setcounter{page}{1}

\title[]{Tunable Magnetism of Transition Metal Nanostructures by Hydrogenated Graphene}

\author{T. Tsevelmaa$^{1,2}$}
\author{Chunfeng$^3$}
\author{D. Odkhuu$^{1,3}$}
\email{odkhuu@inu.ac.kr}
\author{N. Tsogbadrakh$^3$}
\email{Tsogbadrakh@num.edu.mn}
\author{S. C. Hong$^2$}
\affiliation{$^1$Department of Physics, Incheon National University, Incheon 22012, Republic of Korea\\
$^2$Department of Physics and EHSRC, University of Ulsan, Ulsan 680-749, Republic of Korea \\
$^3$Department of Physics, National University of Mongolia, Ulaanbaatar -- 14201, Mongolia}

\begin{abstract}
Controlling magnetism of transition metal atoms by pairing with $\pi$ electronic states of graphene is intriguing. Herein, through first - principle computation we explore the possibility of switching magnetization by forming the tetrahedral $sp^3$ – metallic $d$ hybrid bonds. Graphene multilayers capped by single - layer cobalt atoms can transform into the $sp^3$ - bonded diamond films upon the hydrogenation of the bottom surface. While the conversion is favored by hybridization between the $sp^{3}$ dangling bonds and metallic $d_{z^2}$ states, such a strong hybridization can lead to the reorientation of magnetization easy axis of cobalt adatoms in plane to perpendicular. The further investigations identify that this anisotropic magnetization even can be modulated upon the change in charge carrier density, suggesting the possibility of an electric - field control of magnetization reorientation. These results provide a novel alternative that would represent tailoring magnetism by means of degree of the interlayer hybrid bonds in the layered materials.
\end{abstract}

\pacs{61.48.Gh, 73.22.Pr, 75.30.Gw}

\keywords{DFT, MAE, physisorption, chemisorption, graphene}

\maketitle

\section{Introduction}

The modern field of material science has centered on unique properties of one - to - few atom thick two - dimensional (2D) materials. A prototypical example of one - atom thick 2D system that exhibits a various of fascinating phenomena is graphene \cite{Novoselov, Li}. In particular, the engineering of the chemical and physical properties of graphene by decorating with functional or metallic atoms is the most intriguing \cite{Giovannetti, Odkhuu1}. Conversely, the presence of graphene substantially alters the electronic and magnetic properties of the transition metal atoms, which depends on the degree of hybridization between the metal $d$ orbitals and graphene $\pi$ bands \cite{Krasheninnikov}.

Exploring a thermally stable perpendicular magnetization or magnetic anisotropy (PMA) in otherwise nonmagnetic graphene is at the heart of spintronics research. Yet, there have been a few remarkable studies on the magnetic anisotropy of transition metal atoms, mainly cobalt, on graphene \cite{Xiao, Donati1, Bairagi, Yang}; for example, unexpectedly large PMA up to an order of 100 meV was predicted in cobalt dimer - benzene pairs \cite{Xiao}. On the other hand, individual Co atoms adsorbed onto graphene on a Pt (111) exhibited an in - plane magnetic anisotropy \cite{Donati2}. Interestingly, subsequent experiments have shown that the preferable magnetization axis of the Co adatoms on graphene is the underlying metal substrate dependence: out - of - plane in Ru (0001) and in - plane magnetization in Ir (111) \cite{Donati1}. The authors attributed this magnetization reorientation to the modified hybridization between the Co($3d$) and graphene $p$ bands by the different graphene/metal interactions (chemisorbed graphene/Ru and physisorbed graphene/Ir and /Pt) \cite{Donati1, Donati2}. In more recent studies, through C($p_z$) – Co($d_{z^{2}}$) hybridization, the presence of fullerene molecules reverses magnetization easy axis of the underlying Co films in plane to perpendicular \cite{Bairagi}.

In addition to these remarkable findings, the promising alternative for tailoring the ferromagnet’s anisotropy seemingly resides in the use of even stronger bonding features between the tetrahedral $sp^{3}$ and metallic $d$ orbitals \cite{Odkhuu1}. In this article, we propose such argument where the magnetization easy axis of the freestanding Co (0001) monolayer is reversed from in plane to perpendicular by forming the $sp^{3}$ – $d_{z^{2}}$ hybrid bonds, which is driven by the chemical functionalization of the bottom surface of graphene layers. The further investigations identify that this anisotropic magnetization even can be modulated upon the change in charge carrier density, suggesting the possibility of an electric - field control of magnetization reorientation.

\section{Methodology} 

Density - functional theory (DFT) calculations were performed using the Vienna {\it ab initio} simulation package (VASP) \cite{Kresse}, and exchange - correlation interactions were described with the generalized gradient approximation (GGA) formulated by Perdew, Burke, and Ernzerhof (PBE) \cite{Perdew}. The modeled structure shown in Fig. 1(b) contains a single layer of cobalt atoms deposited on AB - stacked (Bernal - type stacking) bilayer graphene, the bottom surface of which is chemically treated with hydrogen atoms. For a reference, the one - side hydrogenated bilayer graphene is illustrated in Fig. 1(a). An energy cutoff of 400 eV and a 21 x 21 x 1 k - mesh were imposed for the lattice and ionic relaxation, where forces acting on atoms were less than $10^{-2}$ eV/$\mathring{A}$. Spin - orbit coupling (SOC) term is included using the second - variation method employing the scalar - relativistic eigenfunctions of the valence states \cite{Koelling}. Magnetic anisotropy energy (MAE) is obtained based on the total energy difference when the magnetization directions are in the xy - plane ($E^{\parallel}$) and along the z - axis ($E^{\perp}$), $MAE$ = $E^{\parallel}$ - $E^{\perp}$. A dense k - points of 41 x 41 x 1 was used for MAE calculations, which was sufficient to get reliable results.

\section{Results and Discussion}

We first investigated the structural stabilities of the Co/graphene heterostructure under the bottom surface functionalization. Simulating experimental generation of gas phase atoms of hydrogen that can cover up to half the surface of graphene \cite{Luo}, one of every two atoms of the bottom surface of bilayer graphene is chemisorbed to a hydrogen atom, as shown in Fig. 1(a). The three principle adsorption sites of hydrogen on graphene are taken into account so that to define the most stable atomic structure: namely the hollow, bridge, and top sites. The total energy calculations show that the most stable adsorption site of hydrogen atoms is the top site. After the chemisorption of hydrogen atoms, the two graphene layers can be separated by the ’van der Waals’ (vdW) distance or can form interlayer covalent bonds \cite{Odkhuu1}. It has been already indicated in our previous study that the latter structure, i.e., interlayer graphene bonds, is not favored upon the one - side hydrogenation \cite{Odkhuu1}. By contrast, the presence of metal substrate results in the thermodynamically stable $sp^3$ - bonded carbon films over the metal - free hydrogenated graphene layers \cite{Odkhuu1}. This is also the case for the present system where the other surface of bilayer graphene is covered by the monolayer cobalt atoms (See Fig. 1(b)). Similarly, the two - side hydrogenation (or fluorination) of graphene layers can lead to the favorable C - C interlayer bonds (not shown). Furthermore, we would like to note that the transition barrier from graphene layers to $sp^3$ - bonded carbon films on metal substrate upon the functionalization was found to be negligibly small \cite{Odkhuu1}, which is expected for the present system. It was also reported that no energy barrier is required for the physisorption - chemisorption transition of the two - side fluorinated BN multilayers \cite{Zhang}. 

Fig. 2 shows the formation energies ($H_f$), defined as presented in Ref. \cite{Odkhuu1}, of the one - side hydrogenated graphene layers with (filled) and without metal adatoms (unfilled symbol) relative to the pristine bilayer graphene. We also present the $H_f$ of the Co adatoms deposited on the pristine bilayer graphene before hydrogenation in Fig. 2. The results indicate an importance of the presence of metal adatoms in the interlayer formation. We therefore attribute the C - C and C - Co chemical bonds to the saturation of the otherwise unstable $sp^3$ dangling bonds with metal surface states. The driving force for this is the hybridization between the C($sp^3$) and Co($d_{z^2}$) orbitals at the strong chemical Co - C bonds \cite{Odkhuu1, Rajasekaran}. Furthermore, such $sp^3$ - bonded diamond - like carbon structure with metal and functional atoms is estimated to be thermodynamically and structurally stable for the thicknesses of up to eight carbon layers, analogues to that in the functionalized graphene on metal substrate \cite{Odkhuu1}. Similar results were also found for the two - side fluorinated $sp^3$ - bonded BN multilayers \cite{Zhang}.

To better appreciate the strong $sp^3$ – $d_{z^2}$ hybridization, we plot the electronic band structure and density of states (DOS) of the bonded C and Co atoms in Figs. 3(a,b) and 3(c,d) for the physisorbed and chemisorbed Co/graphene, respectively. The electronic features of the pristine Co and graphene layers remain almost unchanged in the physisorption: the majority (minority) spin states of the ferromagnetic atoms are nearly (partially) filled (unfilled), and a band crossing with Dirac cone shape at the Fermi level of graphene (not shown). On the other hand, as seen in Fig. 3, the feature of common peak structures between the Co($d_{z^2}$) and C($p_{z}$) states is apparent throughout the energy level in the chemisorption configuration, indicating the strong orbital hybridization therein. In particular, the existence of these bands in the majority spin state right at the Fermi level is prominent. Such metal -induced gap states (MIGS) can penetrate into up to several - layer thicknesses of graphene, but the C - site induced magnetism (0.08 $\mu_B$) is confined only to the interface layer.

Fig. 4(a) shows the calculated MAE of the physisorbed and chemisorbed Co/graphene. The MAE changes its sign from negative (- 0.62 meV) to positive (0.78 meV) at the physisorption - chemisorption (or $sp^2$ - $sp^3$) transition, which are also well reproduced for the thicker graphene layers. The former and latter stand for the preferable direction of magnetization parallel and normal to the film plane, i.e., PMA. This indicates that the magnetization easy axis of the Co adatoms can be switched and undergoes a transition from an in - plane to perpendicular magnetization upon the formation of C($sp^3$) - Co($d$) hybrid bonds, as schematically illustrated in the inset of Fig. 4(a). We further inspect the relationship between the orbital moment mo and MAE according to Bruno’s model \cite{Bruno}:       

\begin{equation}
MAE = - \frac{\zeta}{4\mu_B}\Delta m_0
\end{equation} 

where $\zeta$ is the strength of SOC and $\Delta m_0 = m^{\parallel}_0 - m^{\perp}_0$. The calculated $\Delta m_0$ of the Co adatoms physisorbed and chemisorbed on graphene are shown at the bottom in Fig. 4(a), where $\Delta m_0 < 0$ and $\Delta m_0 > 0$, respectively. These results adequately obey the Bruno relation: the easy magnetization axis coincides with the direction that has the largest orbital moment.

The $sp^2 - sp^3$ transition evolves in different energy landscapes around the Fermi level, which consequently modulates the MAE. To convince this argument, we follow the recipe of the second - order perturbation theory by Wang {\it et al}. \cite{Wang1}: MAE is determined by the SOC between occupied and unoccupied bands as 

\begin{equation}
MAE = \zeta^2 \sum_{o,u}\frac{|<\Psi_o|L_z|\Psi_u>|^2 - |<\Psi_o|L_x|\Psi_u>|^2}{E_u - E_o}
\end{equation} 

where $\Psi_o$ ($\Psi_u$) and $E_o$ ($E_u$) represent eigenstates and eigenvalues of occupied (unoccupied) states, respectively. Relative contributions of the nonzero   and   matrix elements are 
$<\Psi_{xz}|L_z|\Psi_{yz}> = 1$, $<\Psi_{xy}|L_z|\Psi_{x^2 - y^2}> = 2$, $<\Psi_{z^2}|L_x|\Psi_{xz/yz}> =\sqrt{3}$, $<\Psi_{xy}|L_x|\Psi_{xz/yz}> = 1$ and $<\Psi_{x^2 - y^2}|L_x|\Psi_{xz/yz}> = 1$,
where the positive and negative contributions to MAE are characterized by $L_{z}$ and $L_x$ operators, respectively \cite{Wang1}.
 
In Fig. 4(b), we assign the energy difference of the largest and closest PDOS peaks to the Fermi level in the most relevant orbital states, $d_{xz/yz}$ and $d_{z^2}$, as $E_u - E_o$. The energetics and MIGS were mainly attributed to the $p_z$ – $d_{z^2}$ hybridization in the majority spin state, as addressed in Fig. 3. However, from energy - and k - resolved band analyses, no appreciable coupling of the spin - up occupied and unoccupied $d$ - orbital states appears near the Fermi level. This is in line with the previous full - potential studies on a series of $3d$ - to - $5d$ systems \cite{Odkhuu2, Odkhuu3}, in which the spin channel decomposition terms of MAE that involve the spin - up ($\uparrow$) state, MAE($\uparrow$ $\uparrow$) and MAE($\uparrow$ $\downarrow$), were not significant. For the Co adatoms physisorbed on graphene, there are two strong SOC states between the $d_{z^2}$ and $d_{xz/yz}$ orbitals in the minority - spin state, which leads to the negative MAE through $<\Psi_{z^2}|L_{x}|\Psi_{xz/yz}>$ , where $E_{z^2} - E_{xz/yz} =$ 0.52 and 1.35 eV. When the $sp^2$ transforms to the $sp^3$ phase, the C($sp^3$) -– Co($d$) hybridization splits these minority - spin $d_{z^2}$ states into the low - energy occupied peak at –- 2.3 eV and high - energy unoccupied peak at 1.6 eV. Thus, the negative contributions to MAE decrease; instead, the positive contribution through $<\Psi_{xz}|L_{z}|\Psi_{yz}>$ with $E_{xz} - E_{yz} =$ 0.91 eV becomes more dominant. From the k - resolved MAE and spin - down band analyses, these SOC pairs that involve the $d_{xz/yz}$ and $d_{z^2}$ bands are predominant around the K - M line points, at which the dominant contributions of the MAE are also prominent.

The engineering of the MAE by Fermi level shifts further suggests exploring a crucial effect of the external gating on magnetization reorientation. To anticipate this phenomenon, we analyze the $\Delta m_0$ of the chemisorbed Co/graphene as a function of excess electron per atom, which reflects to the externally injected charge carrier in a positive gating, in Fig. 5(a). Remarkable, the $\Delta m_0$ changes its sign from positive to negative at about 0.1 e/atom. This is due to the nearly flat $d_{z^2}$ band that appears around 0.5 eV above the Fermi level at the K - M points, while the degenerate $d_{xz/yz}$ orbital states become filled. Thereby, the SOC pairs between these filled $d_{xz/yz}$ and empty $d_{z^2}$ bands are formed at the K - M, which should provide the negative MAE(k) therein, as for the case of the physisorption but with the level reversal. As expected, we find that the total MAE changes its sign from the PMA to an in - plane magnetization at around 0.3 e/atom, as seen in Fig. 5(b). These results are of considerable interest in the area of electrically controlled magnetism and magnetoelectric phenomena \cite{Eerenstein, Heron, Wang2}.

\section{ Conclusion }
 
To summarize, our first - principles computation shows that the magnetization easy axis of the monolayer Co (0001) can reorient from in - plane to perpendicular, when the one - side hydrogenated graphene layers are introduced, by forming the $sp^3$ – $d_{z^2}$ hybrid bonds. The chemical functionalization of the bottom surface of graphene layers leads the transformation into thermodynamically stable $sp^3$ - bonded diamond carbon films, which in turn can develop the strong chemical tetrahedral $sp^3$ - metallic $d$ bonds. Moreover, it is found that the perpendicular spin orientation of the Co adatoms chemisorbed onto the $sp^3$ - bonded diamond layers is switchable by altering the density of charge carriers through the application of gate voltage. We thus expect that the present study would provide another novel alternative that would represent tailoring magnetism by means of degree of the interlayer hybrid bonds in the layered materials.

\begin{acknowledgments}
This work is supported by the guest professorship grant (No. P2016-1161) at the National University of Mongolia and the Basic Science Research Program through the NRF funded by the Korean Ministry of Education (NRF-2017R1C1B5017261).
\end{acknowledgments}

\newpage

\begin{figure}
\includegraphics[width=8.0cm]{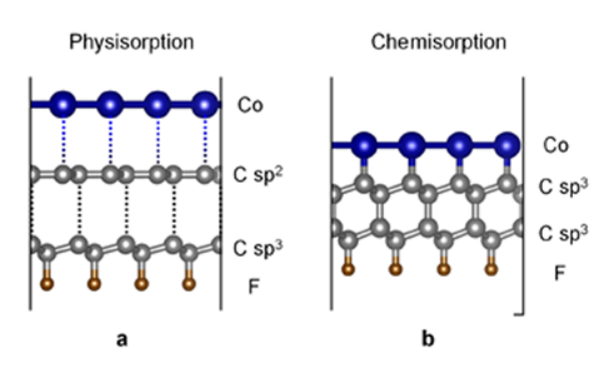}
\caption{(Color online) Side views of the optimized atomic structures for the one - side hydrogenated double layer graphene (a) without and (b) with monoatomic - thick Co adatoms on the other side of graphene surface. The hydrogenation of the outer surface of graphene layers induces the interlayer bonding between the graphene layers. The gray, brown, and blue spheres indicate the C, H, and Co atoms, respectively.}
\label{Fig1}

\includegraphics[width=8.0cm]{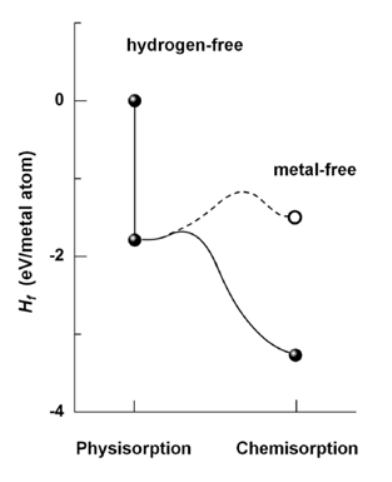}
\caption{The formation energy $H_f$ of the physisorbed and chemisorbed Co/graphene (filled circles). The corresponding result for the chemisorbed bilayer graphene without Co adatoms is shown in open circle (metal - free). Total energy of the Co adatoms on bilayer graphene before functionalization is taken as reference energy (hydrogen - free).}
\label{Fig2}
\end{figure}

\begin{figure}
\includegraphics[width=15.0cm]{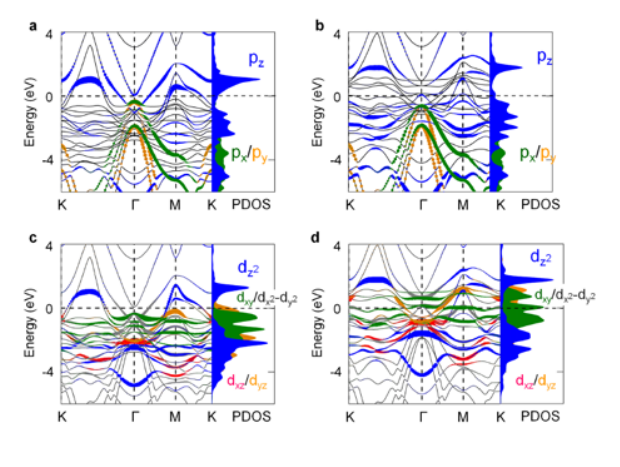}
\caption{(Color online) (a) and (c) Spin - up and (b) and (d) spin - down band structures and density of states of the bonded C and Co atoms of the hydrogenated $sp^3$ graphene layers with Co coverage, respectively. In (a) and (b), the symbols superimposed over the band lines with green, orange, and blue colors represent black, orange, green, red, and blue colors represent the $p_x$, $p_y$, and $p_z$ states of the C atom. In (c) and (d), the green, red, orange, blue and black denote the $d_{xy}$, $d_{xz}$, $d_{yz}$, $d_{z^2}$ and $d_{x^2 - y^2}$ orbitals of the Co adatom. The size of the symbols is proportional to their weights and the Fermi level is set to zero energy.}
\label{Fig3}
\end{figure}

\begin{figure}
\includegraphics[width=15.0cm]{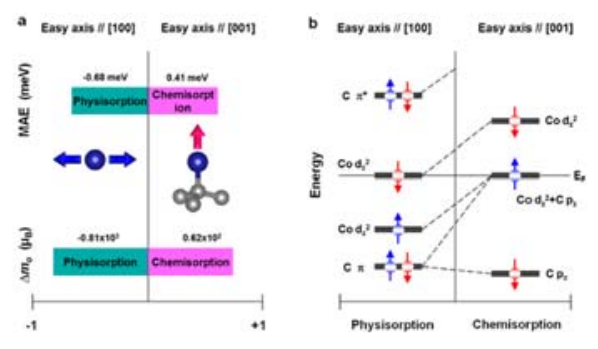}
\caption{(Color online) (a) MAE (upper) and $\Delta m_0$ (lower) of the pristine Co monolayer and Co adatom chemisorbed on the hydrogenated bilayer grahene. (b) Schematic diagram of the single - electron levels of the Co adatom chemisorbed on the hydrogenated bilayer graphene.}
\label{Fig4}

\includegraphics[width=15.0cm]{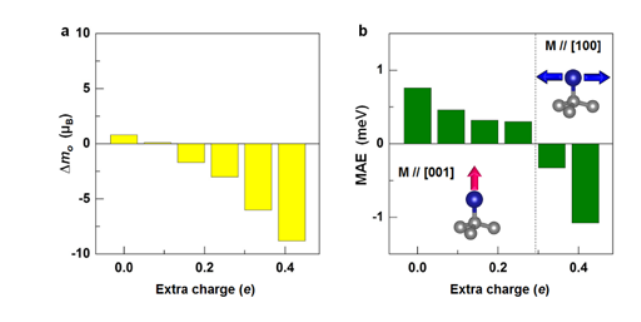}
\caption{(Color online) (a) $\Delta m_0$ and (b) MAE versus the external electron injection of the Co adatom chemisorbed on the hydrogenated bilayer graphene.}
\label{Fig5}
\end{figure}


\begin{thebibliography}{25}
\bibitem{Novoselov} K. S. Novoselov, A. K. Geim, S. V. Morozov, D. Jiang, Y. Zhang, S. V. Dubonos, I. V. Grigorieva, and A. A. Firsov, Science {\bf 306}, 666 (2004).
\bibitem{Li} X. Li, W. Cai, J. An, S. Kim, J. Nah, D. Yang, R. Piner, A. Velamakanni, I. Jung, E. Tutuc, et al., Science {\bf 324}, 1312 (2009).
\bibitem{Giovannetti} G. Giovannetti, P. A. Khomyakov, G. Brocks, V. M. Karpan, J. van den Brink, and P. J. Kelly, Phys. Rev. Lett. {\bf 101}, 026803 (2008).
\bibitem{Odkhuu1} D. Odkhuu, D. Shin, R. S. Ruoff, and N. Park, Sci. Rep. {\bf 3}, 3276 (2013).
\bibitem{Krasheninnikov} A. V. Krasheninnikov, P. O. Lehtinen, A. S. Foster, P. Pyykk¨o, and R. M. Nieminen, Phys. Rev. Lett. {\bf 102}, 126807 (2009).
\bibitem{Xiao} R. Xiao, D. Fritsch, M. D. Kuzmin, K. Koepernik, H. Eschrig, M. Richter, K. Vietze, and G. Seifert, Phys. Rev. Lett. {\bf 103}, 187201 (2009).
\bibitem{Donati1} F. Donati, L. Gragnaniello, A. Cavallin, F. D. Natterer, Q. Dubout, M. Pivetta, F. Patthey, J. Dreiser, C. Piamonteze, S. Rusponi, et al., Phys. Rev. Lett. {\bf 113}, 177201 (2014).
\bibitem{Bairagi} K. Bairagi, A. Bellec, V. Repain, C. Chacon, Y. Girard, Y. Garreau, J. Lagoute, S. Rousset, R. Breitwieser, Y.-C. Hu, et al., Phys. Rev. Lett. {\bf 114}, 247203 (2015).
\bibitem{Yang} H. Yang, A. D. Vu, A. Hallal, N. Rougemaille, J. Coraux, G. Chen, A. K. Schmid, and M. Chshiev, Nano Lett. {\bf 16}, 145 (2016).
\bibitem{Donati2} F. Donati, Q. Dubout, G. Aut`es, F. Patthey, F. Calleja, P. Gambardella, O. V. Yazyev, and H. Brune, Phys. Rev. Lett. {\bf 111}, 236801 (2013).
\bibitem{Rajasekaran} S. Rajasekaran, F. Abild - Pedersen, H. Ogasawara, A. Nilsson, and S. Kaya, Phys. Rev. Lett. {\bf 111}, 085503 (2013).
\bibitem{Kresse} G. Kresse and J. Hafner, Phys. Rev. B {\bf 47}, 558 (1993).
\bibitem{Perdew} J. Perdew, K. Burke, and M. Ernzerhof, Phys. Rev. Lett. {\bf 77}, 3865 (1996).
\bibitem{Koelling} D. D. Koelling and B. N. Harmon, J. Phys. C Solid State {\bf 10}, 3107 (1977).
\bibitem{Luo} Z. Q. Luo and et al., ACS Nano {\bf 3}, 1781 (2009).
\bibitem{Zhang} Z. Zhang, X. C. Zeng, and W. Guo, J. Amer. Chem. Soc. {\bf 133}, 14831 (2011).
\bibitem{Bruno} P. Bruno, Phys. Rev. B {\bf 39}, 865 (1989).
\bibitem{Wang1} D. S. Wang, R. Wu, and A. J. Freeman, Phys. Rev. B {\bf 47}, 14932 (1993).
\bibitem{Odkhuu2} D. Odkhuu, S. H. Rhim, N. Park, K. Nakamura, and S. C. Hong, Phys. Rev. B {\bf 91}, 014437 (2015).
\bibitem{Odkhuu3} D. Odkhuu, S. H. Rhim, N. Park, and S. Hong, Phys. Rev. B {\bf 88}, 184405 (2013).
\bibitem{Eerenstein} W. Eerenstein, N. D. Mathur, and J. F. Scott, Nat. Mater. {\bf 442}, 759 (2006).
\bibitem{Heron} J. T. Heron, M. Trassin, K. Ashraf, M. Gajek, Q. He, S. Y. Yang, D. E. Nikonov, Y.-H. Chu, S. Salahuddin, and R. Ramesh, Phys. Rev. Lett. {\bf 107}, 217202 (2011).
\bibitem{Wang2} W.-G. Wang, M. Li, S. Hageman, and C. L. Chien, Nat. Mater. {\bf 11}, 64 (2012).
\end{thebibliography}
\end{document}